\documentclass[12pt,a4paper]{article}

\usepackage{t1enc}
\usepackage[latin1]{inputenc}
\usepackage[english]{babel}
\usepackage{graphicx}
\begin{document}

\title{A model of population dynamics - \\further investigations}
\author{Iwona Mr\'{o}z \footnotemark[1]\\
Institute of Experimental Physics, University of Wroc{\l}aw\\
Plac Maxa Borna 9, 50--204 Wroc{\l}aw, Poland}
\maketitle
\footnotetext[1]{e-mail: imroz@ifd.uni.wroc.pl}

\begin{abstract}
We present supplementary investigations referring to the model of an evolving population described in 
{\it Physica A 252 (1998) 325--335.} 

The population is composed of individuals characterised by their genetic strings, phenotypes and ages.
We discuss the influence of probabilities of survival of the individuals on the dynamics and phenotypic 
variability of the population.

We show that constant survival probabilities of individuals are propitious for preserving phenotypic variability 
of the population. For constant survival probabilities   
oscillations of 'the average fitness' of the population and normal distributions of the phenotypes are observed.  
When the probabilities of survival are directly proportinal to the individuals' adaptations the population 
can reach the maximum possible average adaptation, but 
the phenotypic variability of the population is completely lost and oscillations of 'the average fitness' of the population 
do not occur. 

We also investigate the behaviour of the population caused by the probabilities 
of survival that 
partly depend on the individuals' adaptations. The role of the length of the individuals' genetic strings is considered 
here.

\end{abstract}

\section{Introduction}
Variability observed in biological populations allows the populations to evolve in different habitats that may in some cases
lead to speciation. Natural populations that have low variability 
are not resistant to changes of their enviromnent and can easily extinct. Preservation of variability is then crucial for 
biological evolution.   
Because of this, it has been investigated intensively by biologists (e.g.\cite{barton}--\cite{doebeli2}),   
and, in recent years, by physicists (e.g.\cite{derrida}--\cite{dasilva}). Variability is considered at different levels: phenotypic 
(the most general), genetic or environmental. Population dynamics is often analysed. 

In biological considerations it is usually assumed that at the phenotypic level the distribution of intensity of phenotypic features 
is normal \cite{slatkin}. In 1996  Doebeli avoided this assumption and presented an interesting model of population dynamics \cite{doebeli}. 
Some of his intriguing conclusions (described below) were tested in 1998 by P\c{e}kalski  \cite{pekalski}, who used a simplified version of the model presented 
in \cite{mroz2}. In this paper we discuss and develop some of P\c{e}kalski's results. 

According to Doebeli's model, a population consists of haploid individuals. Each individual is 
characterised by a genetic 
string that has $c$ genes situated at $c$ loci (a locus is the place at a chromosome where a gene is located). 
Each gene can be in two states: 1 and 0 corresponding 
to two alleles.  
Phenotypes of individuals are characterised by 'a character' $h$ that corresponds to a number of 1's in its 
genetic string. The population is infinite. It can be either sexual or asexual. Generations do not overlap. 
 
In case of a sexual population two individuals can create offsprings. The character of an offspring is established  
by choosing,  
independently for each locus, 
an allele from the alleles of the parents with equal probability. 
The mean fitness of the population, defined as $N(t+1)/N(t)$ (where N(t) denotes 
the total density of the population at time $t$) and the distribution of the phenotypes existing in the population are controlled.   
As a result, it is shown that the mean fitness can oscillate and the type of the oscillations depends on 
the number of loci $c$ . For some parameters, when the total population density is constant, the phenotypes alternate between 
two distributions. Then, the phenotypic variability of the population is preserved and shows unexpected and very 
interesting behaviour. 

Oscillations of the mean fitness of an evolving polulation and strange distribution of phenotypes inspired P\c{e}kalski, who tried 
to confirm Doebeli's results. He used a lattice model based on the standard Monte Carlo simulations. 
According to the model, a population is located on 
a $L \times L$ square lattice. Each  
lattice site may be either empty or contain an individual. The total initial number of individuals is $N(t=0)$. An individual is characterised 
by: its 
location on the lattice $j$, its age $w_j$ and its genome. The individual's age $w_j$ is less or equal to the 
maximum age $M$. $M$ defines the maximum number of Monte Carlo steps (MCS) during which an individual can be 
a member of the population. If the individual's age exceedes $M$, the individual is removed from the lattice. The same 
$M$ is assumed for all individuals. As a parameter, $M$ can vary from 1 to the total duration (in MCS) of a performed simulation.    
The individual's genome is assumed to be a string containing $c$ 
loci with $c$ genes that code $c$ phenotypic features. Genomes and phenotypes are constructed analogically to \cite{doebeli}. 
During the simulation an individual is chosen randomly, its adaptation 
is calculated according to the formula:
\begin{equation}
a_j = {1 \over c} \sum_{k=1}^c f_j^k 
\end{equation}
where $f_j^k$ denotes the fraction of 1's in its genetic string. The individual survives if its adaptation is 
greater than a generated, random number $r\in [0,1]$ (its probability of survival is then strongly connected to the individual's 
adaptation). Then the individual moves across the lattice and meets 
another individual. Movement across the lattice and meeting the neighbour is necessary for mating. 
Adaptation and probability of survival of the neighbour are calculated. If the neighbour 
survives, the individuals mate and create offsprings. The offsprings are located on empty sites inside 
a square $L_G \times L_G$ centered at the first parent location. The number of the offsprings 
depends on the number of empty sites of the square. Their maximum number is $q$. 
The offsprings' phenotypes are established in the same way as described for Doebeli's model.  
After each Monte Carlo step of the simulation the age of all individuals is increased by 1 and all individuals 
which age exceedes the maximum age $M$ are removed from the population. Since individuals 
of different age can mate, generations overlap. 
 
In his paper P\c{e}kalski controlled the time dependence of the density of the population and its average age (relative to the maximum one).
In particular he investigated the average adaptation of the population defined by: 
\begin{equation}
a(t) = {1 \over N(t)} \sum_{j=1}^{N(t)} a_j(t) 
\end{equation}
and the ratio of the numbers of individuals in two succeeding moments of time (Monte Carlo steps). He called this quantity 
the average fitness: 
\begin{equation}
<f(t)> = {N(t+1) \over N(t)}  
\end{equation}

As the main result P\c{e}kalski confirmed Doebeli's conclusion that oscillatory character of the quantity 
$N(t+1)/N(t)$ 
depends on the number of loci
of individuals' genetic strings. The oscillations are damped, their amplitude
depends on $c$ and $M$. The period of oscillations depends on $M$. However, in contrast to Doebeli's results, 
periodic changes of  the distribution of the phenotypes are not observed. It is always normal. Normal distribution of 
the phenotypes indicates that the population contains individuals better and less adapted. The population does not reach 
the maximum possible adaptation, but its phenotypic variability is preserved. The maximum possible adaptation would correspond to the situation when 
the adaptation of every individual equals 1. 
It is suggested that  the population could achieve perfect adaptation, but, as it is shown in Fig.1 of \cite{pekalski} the adaptation of the population 
seems to stabilise at about 0.7. This conclusion is in contrast to the results described in \cite{mroz1} and \cite{mroz2} where initially random populations 
quickly reach the maximum possible adaptation $a=1$ (see e.g. \cite{mroz2}, Fig.2., first region). This fact should be 
considered since, as it has been mentioned above, P\c{e}kalski's model is a simplified version of the model presented in \cite{mroz2}, 
which bases on the model presented in \cite{mroz1}. 
The differences between the models are that in \cite{mroz2}  a population evolves in two different, 
spatially separated habitats and individuals are 
diploids while in \cite{pekalski} a population evolves in one habitat and individuals are haploids. Then it will be interesting to
indicate a reason of such big differences among the adaptations of the populations.

\section{Simulations and Results}
To investigate the conditions under which the adaptation of the population can reach the maximum value and 
other lower values we have performed computer simulations based on the  
model described in \cite{pekalski}. We have used the same parameters as in \cite{pekalski}: a $100 \times 100$ square lattice, 
$L_G=5$, $q=4$ 
and $x(0)=0.3$. Averaging has been done over 25 independent runs. The simulations have been performed for the 
maximum ages $M=5$ and $M=3$ and for the numbers of loci $c=10$, $c=20$ and $c=40$.
  
We have tested populations in which:
\begin{enumerate}
\item Individuals are eliminated from the population only because of aging (when their age is 
greater than the assumed maximum age). In this case 
probability of survival is $p=1.00$.
\item Individuals are eliminated with some constant probability $1-p$, where survival probability 
$p$=0.95; 0.90; 0.85; 0.80. Moreover they are eliminated because of their age.
\item Probability of survival depends on individuals' adaptation, calculated according to the formula (1), 
as assumed in \cite{pekalski}. They are also eliminated because of their age. 
\end{enumerate}

When the population evolves with the probability of survival $p=1.00$, large oscillations of $N(t+1)/N(t)$ are 
observed. The period of the oscillations depends on $M$ (Fig.1), but none of the features of 
the oscillations depends on $c$.  

\begin{figure}
\includegraphics[scale=0.6]{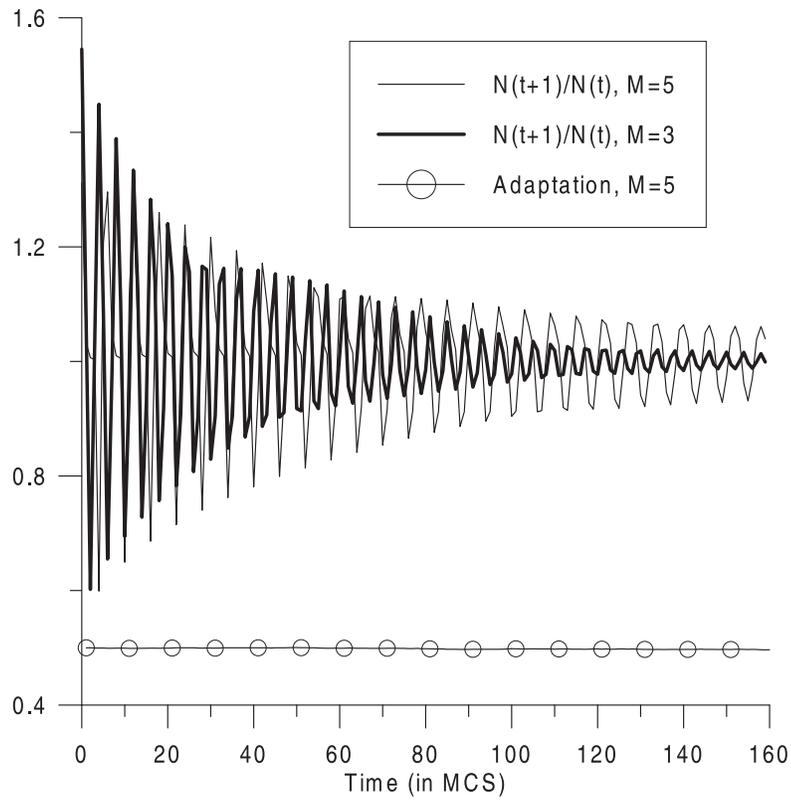}
\caption{Time dependence of $N(t+1)/N(t)$ and the average adaptation of the population for different maximum ages $M$. 
Probability of survival of individuals $p=1.00$. Number of loci $c=20$.}
\end{figure}

This can be explained as follows: before creating offsprings an individual has to move and meet 
a neighbour.  When  the population density is high, the individual can not move. Even if it manages to move and meets 
the neighbour, there is not enough space for many offsprings. Individuals have to be eliminated because of aging, 
then the population density becomes lower, some space required for mating occurs and new offsprings are 
created. At the beginning of the simulations phenotypes are randomly chosen so the average adaptation 
of the population is 0.5. Since there are not many factors that 
may influence the average adaptation (individuals are eliminated only because of their age), 
the population is never adapted well and its average adaptation 
is constant (equals 0.5). 

When individuals are eliminated with some constant probability, independently on their adaptation, 
the average adaptation of the population is also low and constant (still equals 0.5). The oscillations of $N(t+1)/N(t)$ 
are however smaller (Fig.2). The additional mechanism of individuals' elimitation causes that 
there is more free space on the lattice. This results in perturbations of big, age-dependent oscillations.  

\begin{figure}
\includegraphics[scale=0.6]{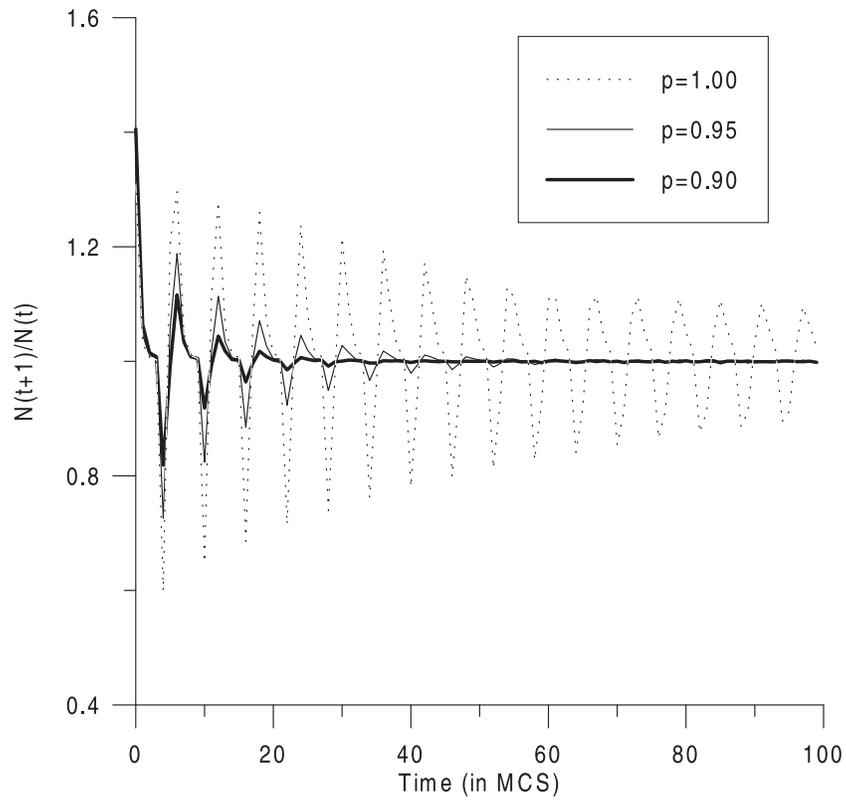}
\caption{Time dependence of $N(t+1)/N(t)$ for different probabilities of survival of individuals. Number 
of loci $c=20$. Maximum age $M=5$.}
\end{figure}

When probability of survival depends on the individual's adaptation, the 
population achieves the maximum possible adaptation and, in contrast to the results presented in 
\cite{pekalski}, it reaches the average adaptation equal 1 independently on the number of loci $c$ (Fig.3). Oscillations 
of $N(t+1)/N(t)$ are not observed here (Fig.4).  When the average adaptation of the population equals 1 all individuals 
are identical - their genetic strings contain only 1's. Then, a typical, normal distribution 
of phenotypes is not observed and the phenotypic variability of the population is lost.

\begin{figure}
\includegraphics[scale=0.6]{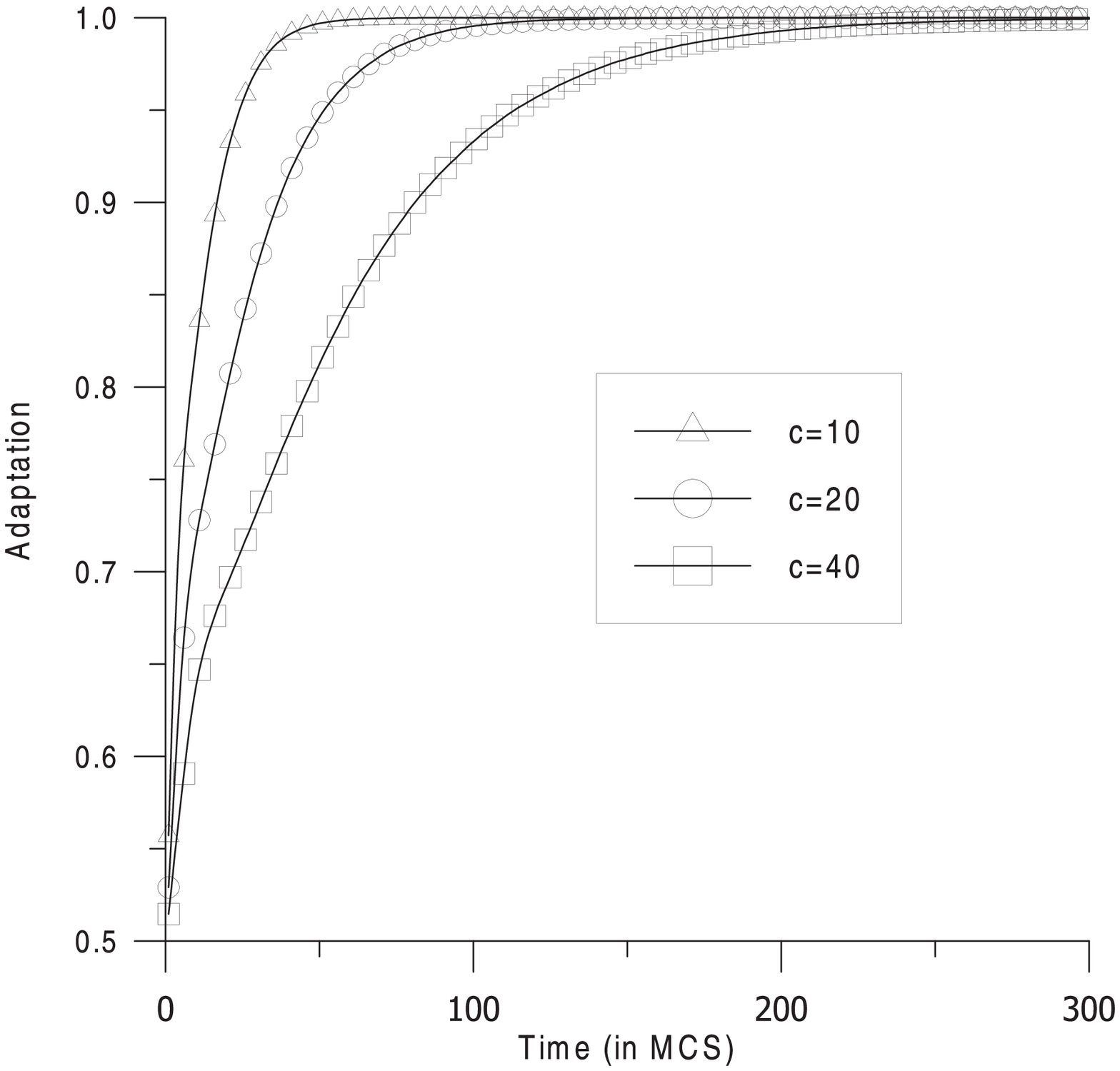}
\caption{Time dependence of the average adaptation of the population when the probability of survival of an individual strongly 
depends on its adaptation. Maximum age $M=5$. The parameter is the number of loci $c$.}
\end{figure}

\begin{figure}
\includegraphics[scale=0.6]{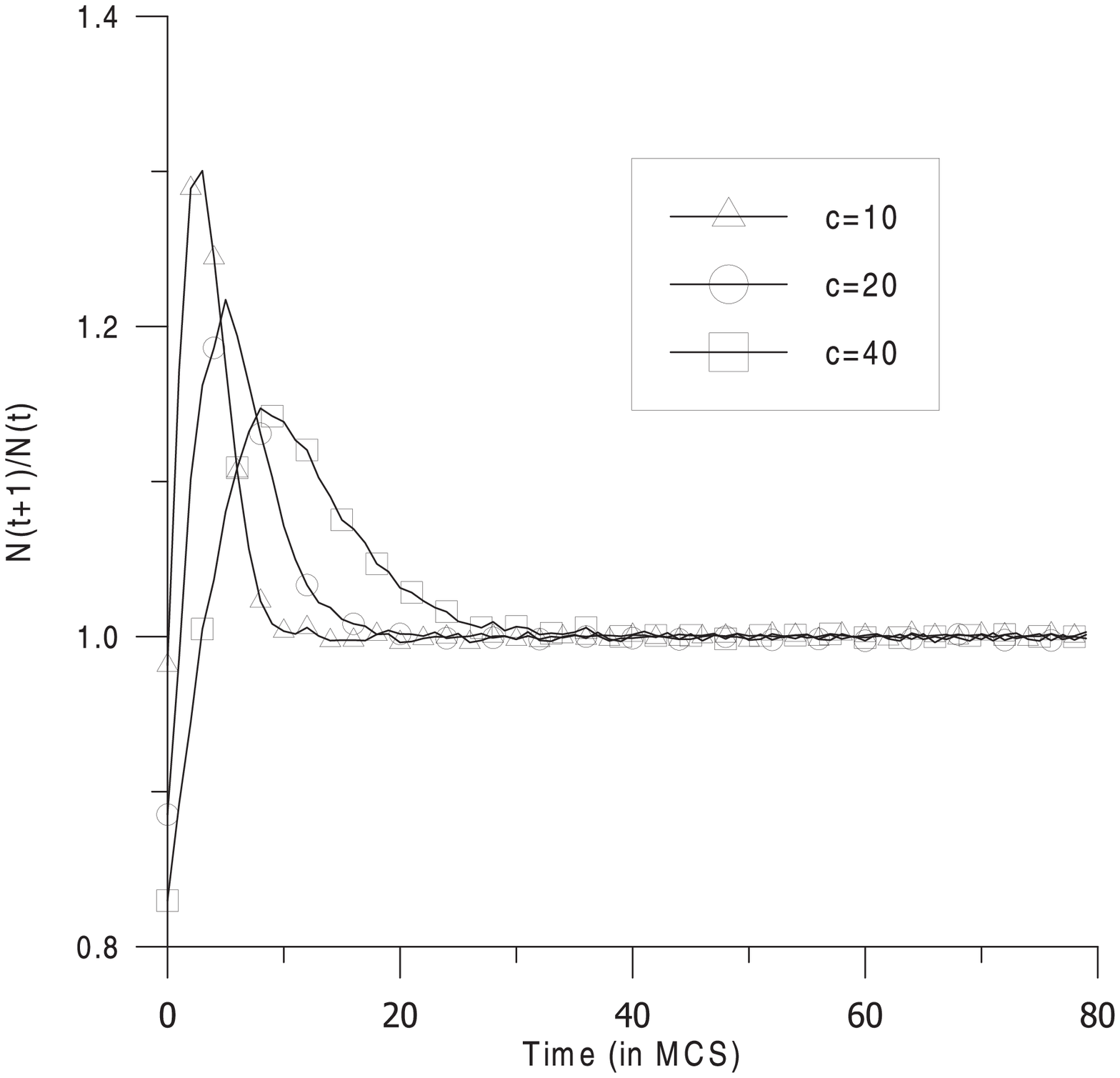}
\caption{Time dependence of $N(t+1)/N(t)$ when the probability of survival of an individual strongly 
depends on its adaptation. Maximum age $M=5$. The parameter is the number of loci $c$.}
\end{figure}
  
The above described procedures lead to two types of populations: a badly adapted one and a perfectly adapted one. 
The average adaptation of 0.7 presented in \cite{pekalski} seems to be an intermediate case.
\begin{figure}
\includegraphics[scale=0.6]{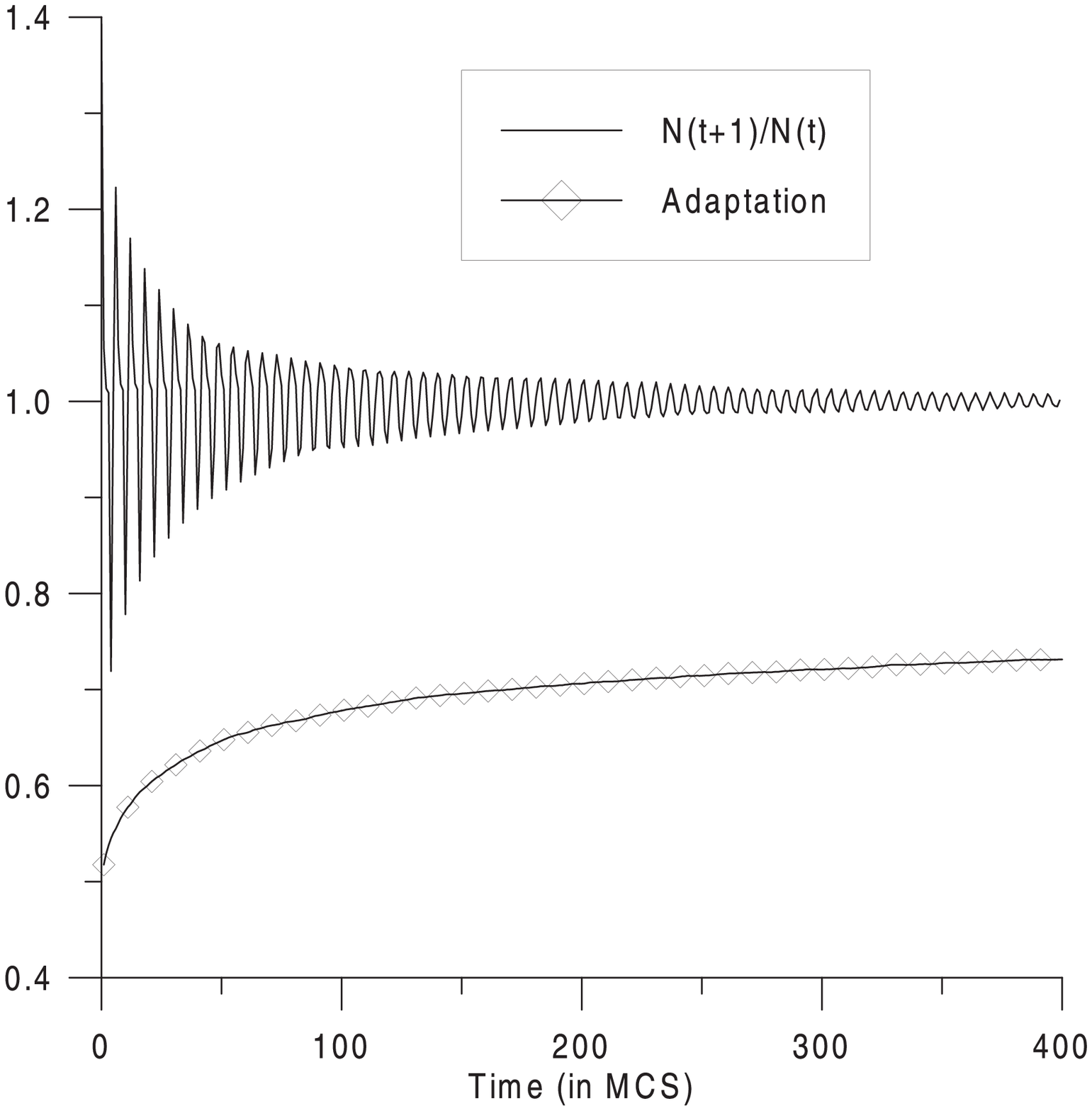}
\caption{Time dependence of $N(t+1)/N(t)$ and the average adaptation of the population. 
Probability of survival of an individual depends on its adaptation according to (4). Number of loci $c=20$, 
$B=10$. Maximum age $M=5$.}
\end{figure}

\begin{figure}
\includegraphics[scale=0.6]{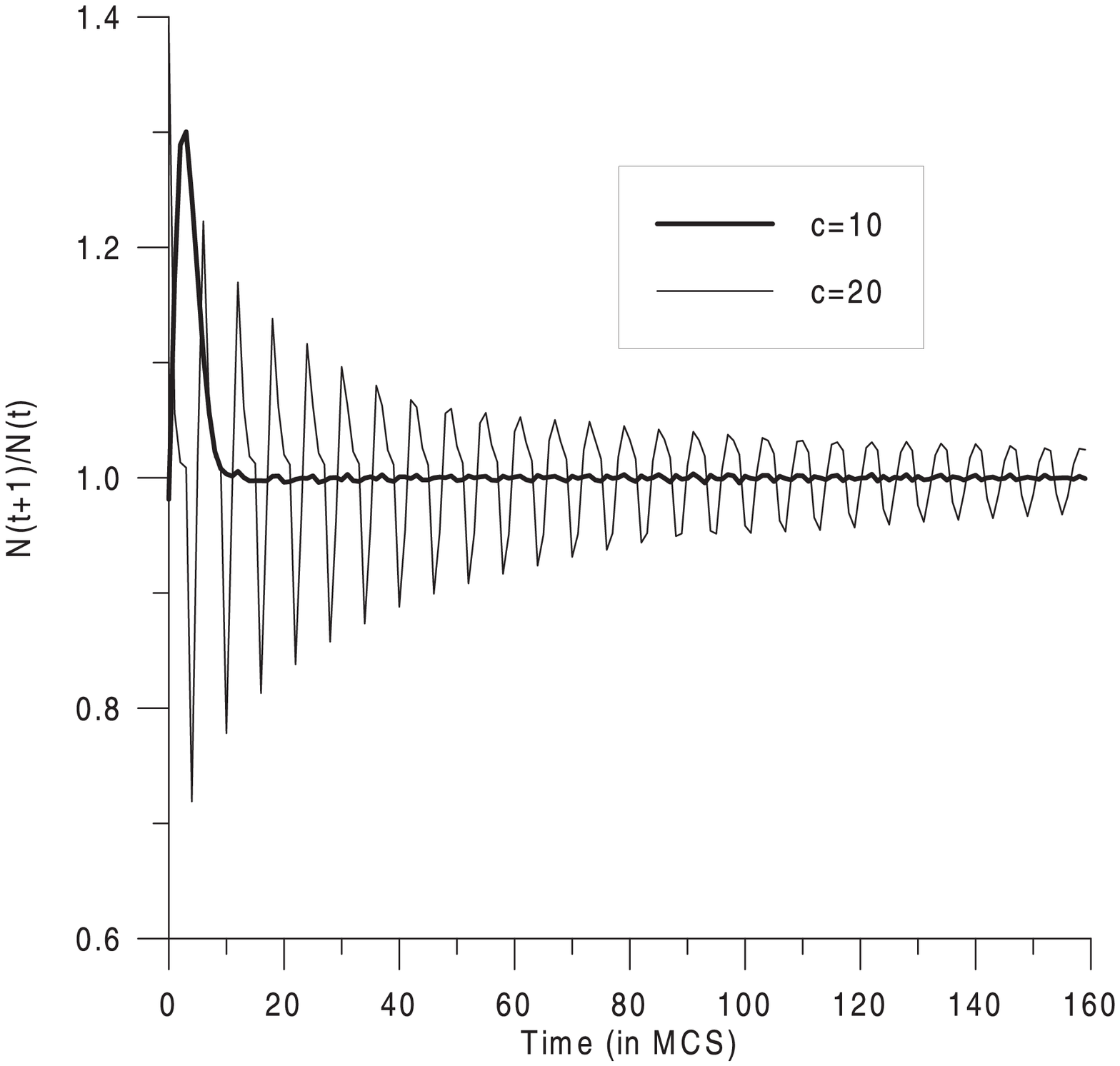}
\caption{
Time dependence of $N(t+1)/N(t)$. Probability of survival of an individual depends on its adaptation according to (4). 
$B=10$. 
Maximum age $M=5$. The parameter is the number of loci $c$.}
\end{figure}

It is 
possible to obtain such an average adaptation if the probability of survival of an individual 
depends on its adaptation, but not so strongly as the probability calculated previously, according to the formula (1). 
For example,  
(1) can be transformed to     
\begin{equation}
A_j = {1 \over B} \sum_{k=1}^c f_j^k 
\end{equation}
where $B$ is a constant equal or smaller than $c$. $A_j$ may be considered as an individual's adaptation, but it must 
be assumed that all values of $A_j$ equal or greater than 1 denote perfect adaptation. For example, let $c=20$ and 
$B=10$. The individuals that have ten or more 1's in their phenotypes will surely survive ($A_j=1$ for ten 1's and $A_j$
is greater than 1 for more than ten 1's. The maximum  possible $A_j=2$ characterises an individual with all 1's in its phenotype). 
Therefore, only really badly adapted individuals can be eliminated. The effect becomes stronger  
for increasing $c$.  At the same time, perfectly adapted individuals, eliminated only because of  their age, may cause oscillations 
of $N(t+1)/N(t)$.

Now it is possible 
to receive oscillating $N(t+1)/N(t)$ and the average adaptation of a population that is between 0.5 and 1. 
The results for $c=20$ and $B=10$ are presented in Fig.5. 

When $B=10$ and $c=10$ the formulas (1) and (4) are identical. Fig.6 presents the time dependence of $N(t+1)/N(t)$ for different $c$. 
Probability of survival is calculated using the formula (4).  
There is some similarity between the results presented in Fig.5 and Fig.6 and, respectively, Fig.1 and Fig.2 of \cite{pekalski}.

\section{Conclusions}
To summarise, for the presented model the oscillatory character of $N(t+1)/N(t)$ and 
values of the average adaptation of the population depend on the way how individuals are eliminated from the 
population.
\begin{enumerate} 
\item{When the individuals are eliminated only because of exceeding the maximum possible age, big, damped oscillations of  
$N(t+1)/N(t)$ are observed while the average adaptation is low. In this case high phenotypic variability (and normal 
distribution of phenotypes) is preserved.} 
\item{The oscillations can be reduced without affecting the phenotypic variability of the 
population 
if the individuals are eliminated with some constant probability.} 
\item{If the individual's survival probability depends on the individual's 
adaptation and it is calculated according to the formula (1), the oscillations of $N(t+1)/N(t)$ do not occur and the population 
quickly reaches perfect adaptation. All individuals are identical and the population has no phenotypic variability.} 
\item{When individuals characterised by the lowest adaptation are eliminated according to the formula (4) it is possible to observe 
many values of the average adaptation of the population and oscillations of $N(t+1)/N(t)$. In this case the population 
can be better adapted than when the indivduals are eliminated because of aging. Moreover, the phenotypic variability 
of the population is preserved. The average adaptation of the population depends on the number of 
individuals' phenotypic features (the number of phenotypic features of an individual equals the number of loci $c$ in its genetic string). 
Then, populations composed of organisms that have different $c$ might evolve in a different way: 
for small $c$ the phenotypic variability would be lost while for bigger $c$ the phenotypic variability would be preserved. 
Therefore, for populations characterised by small $c$ other mechanisms, for example mutations, should be introduced 
to keep the variability. This case seems to be interesting also from biological point of view and we hope that we will 
investigate it in details.}  
\end{enumerate}

\noindent
{\bf Acknowledgements}\\
This work was supported by The State Committee for Scientific Research (grant no. 2PO3B 149 18) and  
University of Wroc{\l}aw (grant no. 2016/ W/ IFD/ 02).\\
I also thank Professor Andrzej P\c{e}kalski (Institute of Theoretical Physics, University of Wroc{\l}aw) for discussion
and Professor Marcel Ausloos (University of Li\`{e}ge) for his valuable comments.

\end{document}